\documentstyle[12pt]{article}

\begin{document}

\begin{center}

{\Large Areas of the Event Horizon and Stationary Limit Surface for a Kerr
Black Hole }

\medskip 

{\bf C. A. Pickett and J. D. Zund}$^{a}$

{\it Departments of Mathematical Sciences and Physics, }

{\it New Mexico State University, Las Cruces, New Mexico, 88003} \\

\bigskip
\end{center}

\noindent {\bf I. INTRODUCTION} 
\medskip

The notions of an event horizon, (EH), and stationary limit surface, (SLS), are important concepts in black hole physics. The former requires no comment, while the latter is less well-known. The SLS geometrically bounds a region outside the EH known as an ergosphere. Inside the ergosphere a test particle cannot travel along the orbit of a timelike Killing vector and remain at rest$^{1}$, and the red shift is infinite on the SLS$^{2}$.

While surface area of an EH has been related to the entropy of a black hole$^{3}$, that of the SLS has not been given a physical interpretation. In this note our approximate evaluation of the area of a SLS suggests a reinterpretation of the area of an EH. 
\bigskip 

\noindent {\bf II. THE KERR BLACK HOLE} 
\medskip

Employing the original form of the uncharged Kerr solution$^{4,5}$, (KS),
expressed in spherical polar coordinates $(\theta ,\phi )$, the angular part 
$d\omega ^{2}$ of the line element $ds^{2}$ is given by 
\begin{equation}
d\omega ^{2}=\rho ^{2}d\theta ^{2}+\sin ^{2}\theta \left[ r^{2}+a^{2}+\frac{%
2mr}{\rho ^{2}}\ a^{2}\sin ^{2}\theta \right] d\phi ^{2}
\end{equation}
where 
\begin{equation}
\rho ^{2}=r^{2}+a^{2}\ \cos {}^{2\;}\theta 
\end{equation}
with $0<\theta <\pi ,\ 0\leq \phi \leq 2\pi $. This expression employs
geometric, or relativistic, units in which the speed of light and the
Newtonian gravitational constant are unity. Quantity $a$ is then the angular
momentum $J$ per unit mass of a spinning particle of mass $m$. Note that $%
a\neq 0$ for the KS, and when $a=0$ the KS reduces to the Schwarzschild
solution, (SS).

Rewriting (1) as

\begin{equation}
d \omega^{2} = \gamma_{AB}\ dx^{A} dx^{B}
\end{equation}
where $x^{A} = (\theta , \phi),\ A = 2, 3$, the area ${\bf A}$ of the corresponding 2-dimensional (angular) surface is given by 
\begin{equation}
{\bf A} = \int_{0}^{2 \pi} \int_{0}^{\pi} \sqrt{\det \gamma_{AB}}\ d\theta d\phi .
\end{equation}

Inspection of $\det \gamma_{AB}$ shows that it is independent of $\phi$, so consequently (4) reduces to

\begin{equation}
{\bf A} = 2\pi \int_{0}^{\pi} \sqrt{\det \gamma_{AB}}\ d\theta.
\end{equation} 
\bigskip

\noindent {\bf III. SURFACE AREAS OF THE EH AND SLS} 
\medskip

By definition, for an EH
\begin{equation}
r = r_{EH} = m + \sqrt{m^{2} - a^{2}}
\end{equation}
and for a SLS
\begin{equation}
r = r_{SLS} = m + \sqrt{m^{2} - a^{2} \cos^{2} \theta}
\end{equation}

Then, upon using these values of $r$ in (5), one obtains expressions for the
respective surface areas ${\bf A}_{EH}$ and ${\bf A}_{SLS}$. \ Of course,
for the SS, since $a=0$ the EH and the SLS coincide.

Evaluation of (5) for the EH is elementary, and one immediately obtains
\begin{equation}
{\bf A}_{EH} = 8 \pi m \left(m + \sqrt{m^{2} - a^{2}} \right).
\end{equation}

However, the evaluation of (5) for the SLS is non-trivial, and involves the integral
\begin{equation}
{\bf A}_{SLS} = 4\pi m \int_{0}^{\pi} r_{SLS} \sqrt{1 + \frac{a^{2}}{mr_{SLS}} \sin^{2}\theta} \cdot \sin\theta\ d\theta.
\end{equation}

The difficulty calculating this is obvious since $r_{SLS}$ involves $\theta$, and appears {\it twice} in the integrand!

To the best of our knowledge, no one has succeeded in evaluating the integral (9) in closed form. It is perhaps noteworthy that in his treatise$^{5}$ Chandrasekhar calculated ${\bf A}_{EH}$ and gave the result (8), but made no comment about the evaluation of (9). Indeed, he did not even cite an approximate expression for ${\bf A}_{SLS}$. 
\bigskip

\noindent {\bf IV. AN APPROXIMATE EVALUATION OF $A_{SLS}$} 
\medskip

Since the expression (9) seems to be intractable it is natural to seek an approximate determination of ${\bf A}_{SLS}$.

Fortunately, both the uncharged KS and the `No Hair' theorem require that 
\begin{equation}
m^{2} - a^{2} \geq 0,
\end{equation}
and this permits us to obtain an approximate expression for ${\bf A}_{SLS}$. The elementary inequality, which is easily proven, 
\begin{equation}
a^{2}/m^{2} \ll 1 \Rightarrow a^{2}/mr_{SLS} \ll 1, 
\end{equation}
then permits us to do a binomial expansion
\begin{equation}
(1\pm X)^{1/2}\simeq 1\pm \frac{1}{2}X,
\end{equation}
with $X \ll 1$, for the radical in $r_{SLS}$.

For ${\bf A}_{SLS}$ this expansion must be used {\it twice}. First, expansion of the radical in the integrand yields
\begin{equation}
{\bf A}_{SLS} \simeq 4\pi m \int_{0}^{\pi} r_{SLS} \left(1 + \frac{a^{2}}{2mr_{SLS}} \sin^{2}\theta\right) \sin\theta \ d\theta
\end{equation}
which we split into two integrals:
\begin{equation}
{\bf A}_{SLS} \simeq 4\pi m \left\{ \int_{0}^{\pi} r_{SLS} \sin\theta\ d\theta + \frac{a^{2}}{2m} \int_{0}^{\pi} \sin^{3}\theta\ d\theta \right\}.
\end{equation}

The first integrand in (14) requires a second expansion,

\[
\normalsize{\begin{array}{lll}
\int_{0}^{\pi} r_{SLS} \sin\theta\ d\theta & \simeq & m \int_{0}^{\pi} \left(1 - \frac{a^{2}}{m^{2}} \cos^{2}\theta \right)^{1/2} \sin\theta\ d\theta \vspace{.1in} \\
& \simeq & m \int_{0}^{\pi} \left(2 - \frac{a^{2}}{2m^{2}} \cos^{2}\theta \right) \sin\theta\ d\theta \vspace{.1in} \\ & = & m \left(4 - \frac{1}{3}\ \frac{a^{2}}{m^{2}} \right);
\end{array}}
\]

while the second integral in (14) is elementary
\[
\frac{a^{2}}{2m^{2}}\int_{0}^{\pi }\sin ^{3}\theta \ d\theta =\frac{2}{3}\ 
\frac{a^{2}}{m^{2}}. 
\]

Adding these two expressions and multiplying by $4\pi m$, we obtain
\begin{equation}
{\bf A}_{SLS} \simeq 16\pi m^{2} + \frac{4\pi}{3} a^{2}
\end{equation}
which is our approximate evaluation of ${\bf A}_{SLS}$. 
\bigskip

\noindent {\bf V. A GEOMETRIC REINTERPRETATION OF $A_{EH}$ AND $A_{SLS}$}
\medskip

Although we have the exact value of ${\bf A}_{EH}$ exhibited in (8), since we have only an approximate value for ${\bf A}_{SLS}$, it seems natural to consider what our approximation procedure gives for ${\bf A}_{EH}$. This is easy, and by using (12) in (8) we obtain 
\begin{equation}
{\bf A}_{EH} \simeq 16\pi m^{2} - 4\pi a^{2}
\end{equation}
which invites a comparison with (15). By recalling that the usual Euclidean surface area of a sphere is
\[
4\pi (radius)^{2},
\]
we observe that
\begin{equation}
16\pi m^{2} = 4\pi(2m)^{2} = 4\pi r_{S}^{2},
\end{equation}
where $r_{S} = 2m$ is the Schwarzschild radius.

This suggests calling the expression in (17), which is the {\it common value} of ${\bf A}_{EH}$ and ${\bf A}_{SLS}$ for the SS, the {\it Schwarzschild area}
\begin{equation}
{\bf A}_{S} = 4\pi r_{S}^{2}
\end{equation}

Of course, for the KS the expressions (15) and (16) are more complicated. However, both the second terms on the right hand sides of these expressions involve the ubiquitous factor of $4\pi$. This suggests introducing the notion of an {\it angular momentum sphere} having $a \neq 0$ as a radius, since dimensionally in geometrized units $a$ is a length. Such a sphere has the Euclidean surface area
\begin{equation}
{\bf A}_{J} = 4\pi a^{2},
\end{equation}
and hence we can rewrite (16) and (15) as
\begin{eqnarray}
{\bf A}_{EH} \simeq {\bf A}_{S} - {\bf A}_{J} \nonumber \\
{\bf A}_{SLS} \simeq {\bf A}_{S} + \frac{1}{3} {\bf A}_{J}.
\end{eqnarray}

These show, as is to be expected, that ${\bf A}_{SLS} > {\bf A}_{EH}$ with ${\bf A}_{SLS} = {\bf A}_{EH}$ only for the SS. The values of ${\bf A}_{EH}$ and ${\bf A}_{SLS}$ given in (20) are surprisingly close in contrast to the usual pictorial illustrations of the ergosphere given in the literature.
\bigskip

\noindent {\bf VI. NUMERICAL EVALUATION OF $A_{EH}$ AND $A_{SLS}$}
\medskip

Since the approximate values of ${\bf A}_{EH}$ and ${\bf A}_{SLS}$ in (20)
are curiously close and tantalizingly simple, one wonders whether this is
accidental, or quite sensible. \ While our use of the binomial expansion
(12) seems sensible, are we justified in omitting the higher order terms?

To answer this question we have done a numerical computation of ${\bf A}_{EH}$ and ${\bf A}_{SLS}$ by using Mathcad$^{6}$. This computes definite integrals by using a Romberg algorithm which accelerates the convergence of a sequence of simple trapezoidal / midpoint approximations of the value of the integral by extrapolating both the sequence of estimates and the widths of the subintervals. Since our goal is to check the accuracy of the expressions given in (20), we will use the values of $a$ and $m$ given by Shapiro and Teukolsky$^{7}$, and let Mathcad choose the method of evaluation with an automatic default setting of 0.001.

Then in geometric units: $m = 1.478 \times 10^{5} cm, J = 4.034 \times 10^{9} cm^{2}$ so that $a = 2.73 \times 10^{4} cm$, and
\[
a/m = 0.185.
\]

Upon direct calculation (15) gives
\[
{\bf A}_{SLS} = 1.101 \times 10^{12} cm^{2},
\]
while Mathcad's evaluation is
\[
{\bf A}_{SLS} = 1.094 \times 10^{12} cm^{2}.
\]

Upon taking the latter to be `exact' this yields an error of $5.797 \times 10^{-3} cm^{2}$, i.e. 0.58\%.

Likewise, upon direct evaluation (16) gives
\[
{\bf A}_{EH} = 1.089 \times 10^{12} cm^{2}
\]
while evalution of the exact expression (8) yields
\[
{\bf A}_{EH}=1.089\times 10^{12}cm^{2}, 
\]
and the error between these is $8.128\times 10^{7}cm^{2}$, i.e. 0.008\%.

This numerical comparison gives us confidence in the accuracy of the approximate value of ${\bf A}_{SLS}$ displayed in equation (15).
\bigskip

\noindent {\bf ACKNOWLEDGMENT}
\medskip

The second author would like to express his thanks to Professor George Burleson for encouraging him to offer a course in black hole physics (Spring 1999), where the topic of this current investigation was assigned as a problem.
\bigskip

\noindent a) Electronic mail: jzund@nmsu.edu
\medskip

\noindent 1. See Pankaj S. Joshi, {\it Global Aspects in Gravitation and Cosmology} (Clarendon Press, Oxford, 1993) pp 77-82; and for a less mathematical discussion, Jean-Pierre Luminet, {\it Black Holes}, (Cambridge University Press, Cambridge, 1992).
\medskip

\noindent 2. Remo Ruffini and John A. Wheeler, ``Relativistic cosmology and space platforms" in {\it The Significance of Space Research for Fundamental Physics} (European Space Research Organization, Neuilly-sur-Seine, 1971) pp 72-99. This is the technical version of the authors popular exposition ``Introducing the black hole," Physics Today, January 1971, 30-41.
\medskip

\noindent 3. Jacob D. Bekenstein, ``Black holes and entropy," Physical Review D {\bf 7}, 2333-2346 (1973);
 and ``Black hole thermodynamics," Physics Today, January 1980, 24-31.
\medskip

\noindent 4. Roy P. Kerr, ``Gravitational field of a spinning mass as an example of algebraically special metrics," Physical Review Letters {\bf 11}, 237-238 (1963) and pp 273-318 of reference 5.
\medskip

\noindent 5. Subrahmanyan Chandrasekhar, {\it The Mathematical Theory of Black Holes} (Clarendon Press, Oxford 1983) p. 317.
\medskip

\noindent 6. Mathcad User's Guide: Mathcad 8 Professional (Mathsoft Inc., Cambridge MA, 1998).
\medskip

\noindent 7. Stuart L. Shapiro and Saul A. Teukolsky, {\it Black Holes, White Dwarfs, and Neutron Stars} (J. Wiley \& Sons, New York, 1983) p. 357.

\end{document}